\documentclass{iopart}
\eqnobysec

\newcommand{\text}[1]{\hbox{#1}}

\renewcommand{\d}{{\rm d}}

\newcommand{\frad}[2]{\displaystyle{\displaystyle#1\over\displaystyle#2}}

\newcommand{\eq}{{\rm eq}}
\newcommand{\TRM}{{\rm TRM}}

\newcommand{\EA}{{\rm EA}}

\newcommand{\PP}{{\cal P}}

\input epsf
\usepackage{graphicx}
\begin{document}

\title{Nonequilibrium critical dynamics of ferromagnetic spin systems}

\author{C Godr\`eche\dag\footnote{godreche@spec.saclay.cea.fr}
and J M Luck\P\footnote{luck@spht.saclay.cea.fr}}

\address{\dag\ Service de Physique de l'\'Etat Condens\'e,
CEA Saclay, 91191 Gif-sur-Yvette cedex, France}

\address{\P\ Service de Physique Th\'eorique\footnote{URA 2306 of CNRS},
CEA Saclay, 91191 Gif-sur-Yvette cedex, France}

\begin{abstract}
We use simple models (the Ising model in one and two dimensions, and the
spherical model in arbitrary dimension) to put to the test some recent ideas
on the slow dynamics of nonequilibrium systems.
In this review the focus
is on the temporal evolution of two-time quantities and on the
violation of the fluctuation-dissipation theorem, with special emphasis
given to nonequilibrium critical dynamics.
\end{abstract}

\section*{Prologue}

The aim of this review is to summarise recent works devoted to the
dynamics of
ferromagnetic spin systems after a quench from infinite temperature
to their critical temperature.

The initial impetus for such an investigation was the desire to
put to the test, on simple models, some recent
ideas on the slow dynamics of nonequilibrium systems
(aging of two-time quantities and violation of the
fluctuation-dissipation theorem).
By simple models we
mean models with no quenched disorder, with, for some of them at least, the
virtue of being solvable. Here we address the case of ferromagnetic spin
systems, such as the Ising model in one and two dimensions, and the
spherical model in arbitrary dimension. Urn models are also simple enough to
serve the same purpose. They are the subject of another review in this
volume \cite{glreview}.

During the course of
this investigation we realised the interest of posing the same questions for
nonequilibrium critical dynamics \cite{gl2D,gl1D}.

\section{The fluctuation-dissipation theorem and its violation}

Consider a generic spin system evolving at constant temperature from a
disordered initial configuration.

Let $s$ and $t$, with $s<t$, be two successive instants of time, and
$\tau =t-s$, their difference.
Denoting by $\sigma (t)$ the spin at time $t$, we
consider the correlation
\[
C(t,s)=\langle \sigma (s)\sigma (t)\rangle ,
\]
and the local response to a time-dependent external magnetic field $H(t)$%
\[
R(t,s)=\frac{\delta \langle \sigma (t)\rangle }{\delta H(s)}.
\]

At equilibrium, that is when the waiting time $s$ is large compared to
the equilibration time $\tau _{\eq}$, these functions are stationary. They
only depend on the time difference $\tau $:
\begin{eqnarray*}
C(s,t) &=&C_{\eq}(\tau ) ,\\
R(t,s) &=&R_{\eq}(\tau ),
\end{eqnarray*}
and are related by the fluctuation-dissipation theorem (for a simple
presentation see e.g. \cite{chandler}):
\[
R_{\eq}(\tau )=-\frac{1}{T}\frac{\d C_{\eq}(\tau )}{\d\tau }.
\]
This situation is typical of the high-temperature regime (e.g. $T>T_{c}$ for
a ferromagnet), where $\tau _{\eq}$ is small.

In experiments or simulations, instead of measuring $R(t,s)$, one considers
the integrated response, i.e., either the
thermoremanent magnetisation of the system at time $t$,
$M_{\mathrm{TRM}}(t,s)$, obtained after
applying a small magnetic field $h$, constant between $t=0$ and $s$;
or the zero-field-cooled magnetisation
$M_{\mathrm{ZFC}}(t,s)$, where now $h$ is
constant between $s$ and $t$.
Defining the reduced integrated response $\rho(t,s) $ by
\[
\rho(t,s) =\frac{T}{h}M(t,s),
\]
we thus have
\begin{eqnarray}
\rho _{\mathrm{TRM}}(t,s) &=&T\int_{0}^{s}\d u\,R(t,u) ,
\label{trm}
\\
\rho _{\mathrm{ZFC}}(t,s) &=&T\int_{s}^{t}\d u\,R(t,u).
\nonumber
\end{eqnarray}

At equilibrium, using the fluctuation-dissipation theorem, we have
\begin{eqnarray*}
\rho _{\mathrm{TRM}}(t,s) &= &\int_{0}^{C(\tau )}\d C=C(\tau ) ,\\
\rho _{\mathrm{ZFC}}(t,s) &= &\int_{C(\tau )}^{1}\d C=1-C(\tau ),
\end{eqnarray*}
thus a plot of $\rho $ against $C$ is given by a straight line of slope $+1$
($\rho _{\mathrm{TRM}}$) or $-1$ ($\rho _{\mathrm{ZFC}}$), as soon as $s$ is
large enough.

At low temperature (below $T_{c}$ for a ferromagnet), $\tau _{\eq}$ is either very
large or infinite. In the scaling regime where $%
1\ll s\sim t\ll \tau _{\eq}$, aging takes place, i.e., $C$ and $R$ are no
longer stationary, and the fluctuation-dissipation theorem does not hold.
The question is therefore to determine the relationship between $C$ and $R$,
if any.
This can be done by defining the
fluctuation-dissipation ratio $X(t,s)$ by \cite{ck94,ckpa,revue}
\[
R(t,s)=\frac{X(t,s)}{T}\frac{\partial C(t,s)}{\partial s}.
\]

Assume that, in the scaling regime, all the time dependence of $R$ can be
parameterised by $C$. Or, in other words, that $C$ acts as a clock for $R$.
That is, for $1\ll s\sim t$,
\begin{equation}
X(t,s)\approx X(C(t,s)).
\label{XC}
\end{equation}
As a consequence, we have
\begin{eqnarray*}
\rho _{\mathrm{TRM}}(t,s) &\approx &\int_{0}^{C(t,s)}\d C\,X(C) ,\\
\rho _{\mathrm{ZFC}}(t,s) &\approx &\int_{C(t,s)}^{1}\d C\,X(C).
\end{eqnarray*}
Hence, in a plot of $\rho $ against $C$, the slope at a given point is given
by $\pm X(C)$.

This behaviour has been observed in a number of instances.
In particular, a
census of the different cases of spin systems hitherto studied shows the
existence of three main types of behaviour at low temperature (for a
summary, see \cite{cug}, and references therein).
For domain-growth models, $X(C)$
is discontinuous in $C$, taking a first value equal to~$1$, and a
second one equal to zero \cite{ckpe,barrat,berthier}
(see discussion in section 2).
For spin-glass models with $p$-spin interactions,
$X(C)$ is still discontinuous but the second value is non-zero.
Finally, for continuous spin-glass models, $X(C)$ is a non-trivial curve \cite
{ck94}.

In the present review we show that, at $T=T_c$,
non-trivial statements can be
formulated on the same issue.
Hereafter we specialise to ferromagnetic spin systems.
We take
as representatives the Ising model in one and two
dimensions, and the spherical model in arbitrary dimension.
The Hamiltonian
describing these models reads
\[
E(t)=-J\sum_{(i,j)}\sigma _{i}(t)\sigma _{j}(t)-\sum_{i}H_{i}(t)\sigma
_{i}(t),
\]
where the first sum runs over pairs of neighbouring sites.

For the Ising model, $\sigma _{i}=\pm 1$, and the (non-conserved Glauber)
dynamics is governed by the heat-bath rule:
\[
\PP (\sigma _{i}(t+\d t)=\pm 1)
=\frac{1}{2}\left( 1\pm\tanh \beta h_{i}(t)\right) ,
\]
where the local field reads $h_{i}=\sum_{j}\sigma _{j}+H_{i}$, the sum
running over the neighbours of site $i$.

For the spherical model, $\sigma _{i}$ is a real number
with the constraint $\sum_{i}\sigma _{i}^{2}=N$,
where $N$ is the number of spins
\cite{berlin,stanley,baxter}.
The dynamics is governed by the Langevin
equation \cite{cd}
\[
\frac{\d\sigma _{i}}{\d t}=-\frac{\partial E}{\partial \sigma _{i}}-\lambda
(t)\sigma _{i}+\eta _{i}(t).
\]
In the right side, $\lambda (t)$ is a Lagrange multiplier ensuring the
constraint, and $\eta _{i}(t)$ is a Gaussian white noise with correlation
$\left\langle \eta _{i}(t)\eta _{j}(t^{\prime })\right\rangle =\delta
_{ij}\delta (t-t^{\prime })$.

In both cases, at time $t=0$, the system is in a disordered initial
configuration (e.g. corresponding to equilibrium at infinite temperature).

\section{{\ Aging below }$T_{c}$: {low-temperature coarsening}}

We first describe in more detail the behaviour of correlation and response at
low temperature, for a generic ferromagnetic model such as the spherical
model or the 2D Ising model,
evolving at constant temperature after a quench
from $T=\infty $ down to $T<T_{c}$.
We defer the discussion of the 1D Ising model
to section 4.

In such a situation, domains of opposite
sign grow, with a characteristic size $L(t)\sim t^{1/z}$, where
$z=2$ is the growth exponent \cite{langer,bray}.

In a first regime ($1\sim \tau \ll s$), dynamics is stationary.
Correlations
decay from $C(s,s)=1$, to the plateau value
\[
q_{\EA}=\lim_{\tau \rightarrow \infty }\lim_{s\rightarrow \infty }C(s+\tau
,s)=M_{\eq}^{2},
\]
where $M_{\eq}$ is the equilibrium magnetisation.
Though the system becomes
stationary, it is still coarsening, and therefore does not reach thermal
equilibrium.
However the fluctuation-dissipation theorem holds, and $X=1$.

In the scaling regime where $s$ and $t$ are simultaneously large
($1\ll s\sim t$), with arbitrary ratio $x=t/s$, aging takes place, and
correlations behave as \cite{bray}
\begin{equation}
C(t,s)\approx M_{\eq}^{2}\,f_{C}\left( \frac{t}{s}\right) .
\label{Cts_coars}
\end{equation}
For small temporal separations ($\tau \ll s$, or $x\to1$), we have
$f_C(x)\to1$, implying $C(t,s)\to M_{\eq}^{2}$.
In other words, equation (\ref{Cts_coars}) describes the departure
from the plateau value $M_{\eq}^{2}$.
For well-separated times ($1\ll s\ll t$, or $x\gg 1$)
$f_{C}(x)$ decays algebraically as
\[
f_{C}(x)\approx A_{C}\,x^{-\lambda /z},
\]
where $\lambda $ is the autocorrelation exponent \cite{fisher}. As a
consequence, we have
\[
\frac{\partial C(t,s)}{\partial s}\approx \frac{M_{\eq}^{2}}{s}\,f_{C^{\prime
}}\left( \frac{t}{s}\right) ,
\]
with $f_{C^{\prime }}(x)\approx A_{C^{\prime }}\,x^{-\lambda /z}$, at large $%
x$.

In the same regime it is reasonable to make the scaling assumption
(see discussion below)
\begin{equation}
R(t,s)\approx s^{-1-a}\,f_{R}\left( \frac{t}{s}\right) ,
\label{Rts_coars}
\end{equation}
with an unknown exponent $a>0$, and
with again the decay at large $x$
\begin{equation}
f_{R}(x)\approx A_{R}\,x^{-\lambda /z}.
\label{fR_coars}
\end{equation}
We have therefore
\[
X(t,s)\approx \frac{s^{-a}}{M_{\eq}^{2}}T\frac{f_{R}(t/s)}{f_{C^{\prime
}}(t/s)}\approx \frac{s^{-a}}{M_{\eq}^{2}}T\frac{A_{R}}{A_{C^{\prime }}}.
\]
The fluctuation-dissipation ratio thus vanishes
in the scaling regime, irrespective of the ratio $t/s$.

For instance, for the spherical model, the equilibrium magnetisation reads
\[
M_{\eq}^{2}=1-\frac{T}{T_{c}}
\]
and the correlation $C(t,s)$ is given by (\ref{Cts_coars}) with
\[
f_{C}(x)=\left( \frac{4x}{(x+1)^{2}}\right) ^{D/4},
\]
hence the autocorrelation exponent $\lambda =D/2$.
The response is given, in the scaling regime, by \cite{horner}
\begin{equation}
R(t,s)\approx (4\pi )^{-D/2}\left( \frac{t}{s}\right) ^{D/4}(t-s)^{-D/2},
\label{Rts_coars_sph}
\end{equation}
which is in agreement with the form (\ref{Rts_coars}),
with scaling function
\[
f_{R}(x)=(4\pi )^{-D/2}x^{D/4}(x-1)^{-D/2},
\]
and the exponent $a=D/2-1$.

For the 2D Ising model, the exponent
$\lambda \approx 1.25$ \cite{fisher} is only known numerically.
This is also the case of
the scaling functions $f_C$ and $f_R$~\cite{henkel}.
The latter work is compatible with $a=1/2$,
as predicted in Refs.~\cite{berthier,bray2},
where it is argued that the integrated response
scales as $\rho (t,s)\sim L(s)^{-1}g\left( L(t)/L(s)\right)$
for soft spin models with non-conserved dynamics.

In summary, for a ferromagnetic spin system \cite{barrat,cd,horner,berthier},

\begin{description}
\item [$\bullet $] for short times ($\tau \ll s$), such that
$C(t,s)>M_{\eq}^{2}$, the fluctuation-dissipation theorem holds, and $X=1$;

\item [$\bullet $] for long times ($\tau\sim s$),
such that $C(t,s)<M_{\eq}^{2}$,
the fluctuation-dissipation theorem does not hold, and
$X(t,s)\rightarrow 0$
independently of the ratio $t/s$.

\end{description}

Note that we have
\[
\frac{\d X(C)}{\d C}=\delta \left( C-M_{\eq}^{2}\right) ,
\]
in agreement with the static interpretation of $X(C)$ in terms of
the distribution of overlaps $P(q)$ \cite{franz}.

\section{Aging at $T_{c}$: critical coarsening }

The system is now quenched from $T=\infty $ down to $T_{c}$.

In such circumstances, spatial correlations develop in the system, just as in
the critical state, but only over a length scale which grows like
$t^{1/z_{c}}$, where $z_{c}$ is the dynamic critical exponent.
On scales
smaller than $t^{1/z_{c}}$ the system looks critical, while on larger scales
the system is still disordered.
For instance, the equal-time correlation function
$C_{r}(t)=\left\langle \sigma _{0}(t)\sigma _{r}(t)\right\rangle $ scales as
\[
C_{r}(t)\approx |r|^{-2\beta /\nu }\,g\left( \frac{r}{t^{1/z_{c}}}\right) ,
\]
where $\beta $ and $\nu $ are the usual static exponents.
(A summary of the values of exponents is given in the Table.)
The scaling
function $g(y)$ goes to a constant as $y\rightarrow 0$, while it falls off
very rapidly when $y\rightarrow \infty $.

The same temporal regimes, as defined in the previous section, are to be
considered.
However, their physical interpretation is slightly different,
since the order parameter $M_{\eq}^{2}$ vanishes and symmetry between the
phases is restored.

In the first regime ($ \tau \ll s$), the system again becomes
stationary, so that the fluctuation-dissipation holds.

In the scaling regime ($\tau \sim s$), temporal correlations behave as%
\footnote{%
For simplicity we use the same notation $f_{C}$, $f_{R}$, etc. for the
scaling functions appearing in this section, though they are different from
those appearing in the previous section.
We use the same convention for the
amplitudes $A_{C}$, $A_{R}$, etc.}
\begin{equation}
C(t,s)\approx s^{-a_c}\,f_{C}\left( \frac{t}{s}\right) ,\qquad a_c=2\beta /\nu
z_{c}=(D-2+\eta )/z_{c}.
\label{Cts_crit}
\end{equation}
It is instructive to relate this behaviour to that observed for $T<T_{c}$,
namely, $C(t,s)\approx M_{\eq}^{2}\,f_{C}\left( t/s\right) $. The passage
from one formula to the other one is done by noticing that in the critical
region one has $M_{\eq}\sim |T-T_{c}|^{\beta }\sim \xi _{\eq}^{-\beta /\nu }$.
Replacing $\xi _{\eq}$ by $s^{1/z_{c}}$ implies the replacement of $%
M_{\eq}^{2} $ by $s^{-2\beta /\nu z_{c}}\sim s^{-(D-2+\eta )/z_{c}}$.

At large time separations ($x\gg 1$) we have (see \cite{janssen} for a
derivation in the case of the so-called model A \cite{hohenberg})
\[
f_{C}(x)\approx A_{C}\,x^{-\lambda _{c}/z_{c}},
\]
where $\lambda _{c}$ is the critical autocorrelation exponent \cite{huse},
related to the initial-slip critical exponent $\Theta _{c}$ \cite{janssen}
by $\lambda _{c}=D-z_{c}\Theta _{c}$.

\begin{figure}[htbp]
\begin{center}
\includegraphics[width=.7\linewidth,angle=90]{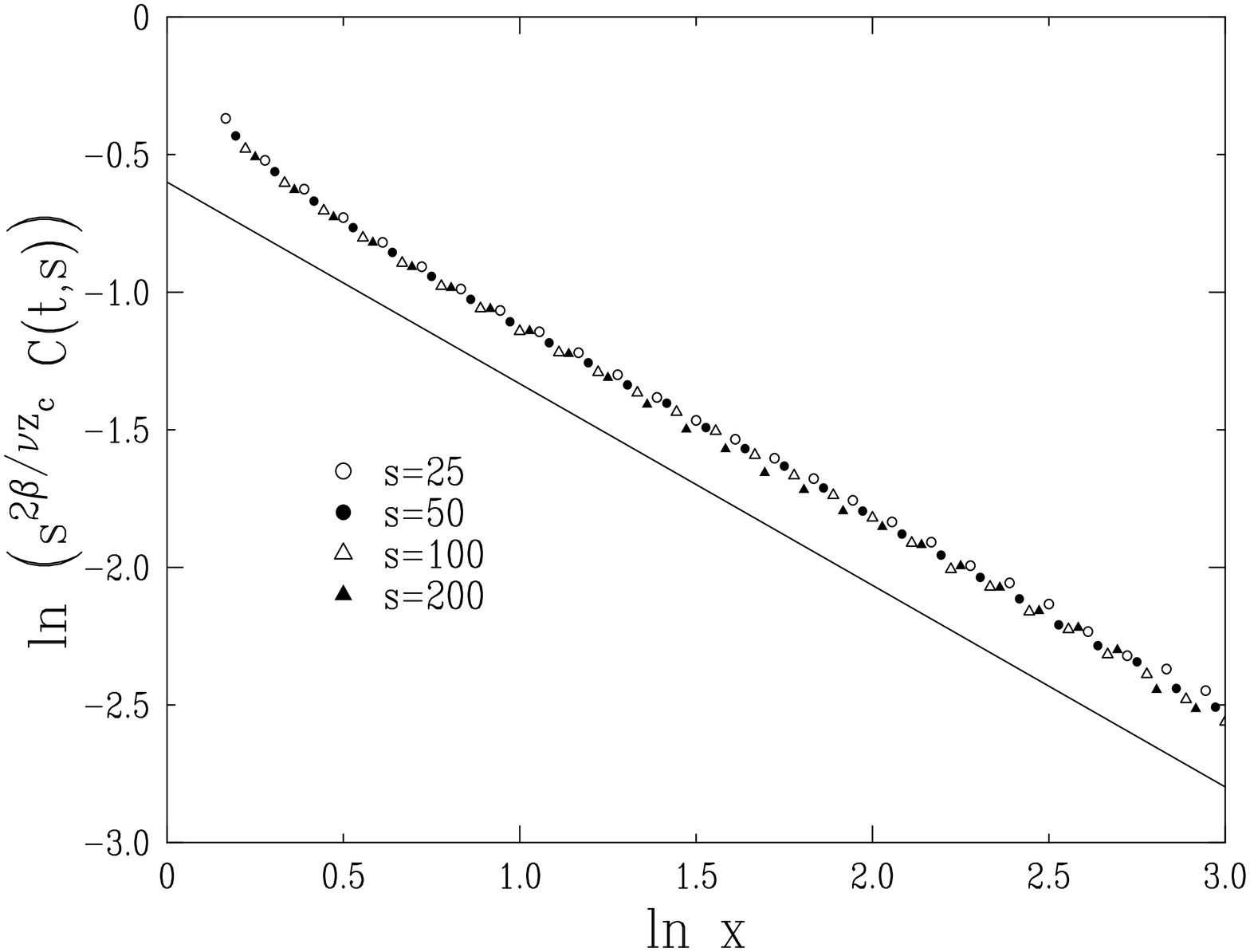}
\caption{Log-log plot of the critical autocorrelation function $C(t,s)$
of the two-di\-men\-si\-o\-nal Ising model,
against time ratio $x=t/s$, for several values of the waiting time~$s$.
Data are multiplied by $s^{2\beta/\nu z_c}$,
in order to demonstrate collapse into the scaling function $f_C(x)$
of eq.~(\ref{Cts_crit}).
Straight line: exponent $-\lambda_c/z_c\approx-0.73$
of the fall-off at large~$x$. (After ref. \cite{gl2D}.)}
\end{center}
\end{figure}

As a consequence of (\ref{Cts_crit}), we have
\[
\frac{\partial C(t,s)}{\partial s}\approx s^{-1-a_c}\,f_{C^{\prime }}
\left( \frac{t}{s}\right) ,
\]
with the decay
$f_{C^{\prime }}(x)\approx A_{C^{\prime }}\,x^{-\lambda_{c}/z_{c}}$
at large $x$.

In the scaling regime, the response function behaves as
\begin{equation}
R(t,s)\approx s^{-1-a_c}\,f_{R}\left( \frac{t}{s}\right) ,
\label{Rts_crit}
\end{equation}
and, for large temporal separations,
\begin{equation}
f_{R}(x)\approx A_{R}\,x^{-\lambda _{c}/z_{c}}. \label{fR_crit}
\end{equation}
(See \cite{janssen} for a derivation of (\ref{Rts_crit}) and (\ref{fR_crit})
in the case of model A.)
Note the similarity of (\ref{Rts_crit}) and (\ref
{fR_crit}) with (\ref{Rts_coars}) and (\ref{fR_coars}), respectively.
The scaling form (\ref{Rts_crit}) of the response implies
\begin{equation*}
\rho _{\mathrm{TRM}}(t,s)
\approx s^{-a_c}\,f_{\rho }\left( \frac{t}{s}\right) ,
\end{equation*}
with, as $x\gg1$, $f_\rho(x)\approx A_\rho x^{-\lambda _{c}/z_{c}}$.

We finally obtain the fluctuation-dissipation ratio
\[
X(t,s)\approx T_{c}\frac{f_{R}(t/s)}{f_{C^{\prime }}(t/s)}={\cal X}%
\left ( \frac{t}{s}\right ),
\]
and,
at large temporal separations,
\[
X_{\infty }
=\lim_{s\rightarrow \infty }\lim_{t\rightarrow \infty}X(t,s)
=\lim_{x\rightarrow \infty }{\cal X}(x)
=T_{c}\frac{A_{R}}{A_{C^{\prime }}}
=T_{c}\frac{A_\rho}{A_C}.
\]
The last equality is equivalent to saying that, for $1\ll s\ll t$,
\begin{equation*}
\rho_{\mathrm{TRM}} (t,s)\approx X_\infty C(t,s).
\end{equation*}
The limit fluctuation-dissipation ratio $X_\infty $ can
thus be measured as the slope near the origin
of the $C-\rho_{\mathrm{TRM}}$ plot.
The scaling function ${\cal X}(x)$,
and in particular the amplitude ratio $X_\infty$,
are universal, in the sense that they neither depend on
initial conditions nor on the details of the dynamics \cite{gl2D,gl1D}.

In the scaling regime, neither $\rho_{\mathrm{TRM}} (t,s)$ nor $X(t,s)$
are functions of $C(t,s)$.
Instead, $X(t,s)$ and $s^{a_c}\rho(t,s)$ are functions of $x=t/s$,
which is in contrast with the situations where equation (\ref{XC}) holds,
and further described in section 1.

We now illustrate the results presented above.
For the spherical model (see the Table for the value of exponents),
the two-time correlation function reads
$$
C(t,s)\approx s^{-(D/2-1)}f_C(x),
$$
where
\[
{\hskip -13pt}f_{C}(x)=\left\{
\begin{array}{ll}
T_c\frad{4(4\pi)^{-D/2}}{(D-2)(x+1)}x^{1-D/4}(x-1)^{1-D/2} & 2<D<4 ,\\ \\
T_c\frad{2(4\pi)^{-D/2}}{D-2}\left( (x-1)^{1-D/2}-(x+1)^{1-D/2}\right) & D>4.
\end{array}
\right.
\]
Thus
\[
\lambda _{c}=\left\{
\begin{array}{ll}
3D/2-2 & 2<D<4, \\
D & D>4.
\end{array}
\right.
\]
Similarly,
the response function behaves as
$$
R(t,s)\approx s^{-D/2}\,f_R(x),
$$
where the scaling function $f_R(x)$ reads
\[
f_{R}(x)=\left\{
\begin{array}{ll}
(4\pi)^{-D/2}x^{1-D/4}(x-1)^{-D/2}& 2<D<4 ,\\ \\
(4\pi)^{-D/2}(x-1)^{-D/2}
& D>4.
\end{array}
\right.
\]
Finally
\[
X_{\infty }=\left\{
\begin{array}{ll}
1-2/D & 2<D<4 ,\\
1/2 & D>4.
\end{array}
\right.
\]

\begin{figure}[htbp]
\begin{center}
\includegraphics[width=.7\linewidth,angle=90]{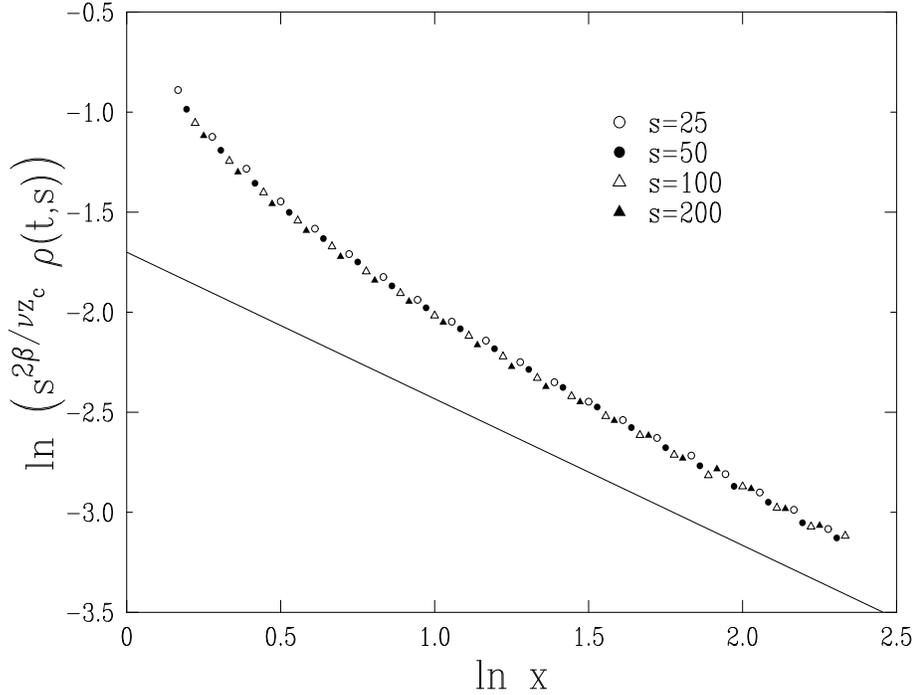}
\caption{Log-log plot of the critical integrated response function
$\rho_{\TRM}(t,s)$ of the two-dimensional Ising model,
against time ratio $x=t/s$, for several values of the waiting time~$s$.
Data are multiplied by $s^{2\beta/\nu z_c}$,
in order to demonstrate collapse into the scaling function $f_\rho(x)$.
Straight line: exponent $-\lambda_c/z_c\approx-0.73$
of the fall-off at large~$x$. (After ref. \cite{gl2D}.) }
\end{center}
\end{figure}

For the 2D Ising model $\lambda _{c}\approx 1.59$ \cite{huse}
is only known numerically.
Figures
1 and 2 show numerical determinations of the scaling functions
$f_{C}$ and $f_{\rho }$ \cite{gl2D}.
In two dimensions, we have $X_{\infty }\approx 0.26$,
and a preliminary study leads to $X_{\infty }\approx0.40$
in three dimensions \cite{gl2D}.

The above discussion can be summarised as follows:

\begin{description}

\item $\bullet$ For short times ($\tau \ll s$), such that
$C(t,s)\gg s^{-2\beta /\nu z_{c}}$, the fluctuation-dissipation theorem holds,
and $X=1$.

\item $\bullet$ For long times ($\tau\sim s$),
such that $C(t,s)\sim s^{-2\beta /\nu z_{c}}$,
the fluctuation-dissipation theorem does not hold.
The fluctuation-dissipation ratio
$X(t,s)$ is given by the scaling function ${\cal X}(t/s)$,
such that ${\cal X}(x)\rightarrow X_{\infty }$ as $x\to\infty$.

\end{description}

This is the critical counterpart of the behaviour of $X(t,s)=X(C)$
for $T<T_{c}$, summarised at the end of section 2.

A last comment is in order.
At thermal equilibrium, for a ferromagnetic system at criticality, the
relationship between magnetic field and magnetisation,
$h\sim M_{\eq}^{\delta}$, is nonlinear.
Therefore linear-response theory, used above to extract
the response of the system, only holds for a magnetic field small
compared to the scale
$h_{0}\sim s^{-\beta \delta /\nu z_{c}}\sim s^{-(D+2-\eta)/2z_c}$.

\section{\ One-dimensional Ising model at $T=0$}

The one-dimensional Ising model is special in the sense that its critical
temperature $T_{c}$ is zero.
Hence the low-temperature phase does not exist.

Another peculiarity of the model stems from the fact that
the magnetisation exponent $\beta$ is equal to zero.
As a consequence, at criticality (i.e., at $T=0$),
there is no temporal prefactor in the expression of $C(t,s)$
(or equivalently, no spatial prefactor in that of $C_r(t)$).
Indeed, let us recall that, at criticality, for
a generic ferromagnetic model, we had
\begin{eqnarray*}
C_{r}(t) &\approx &|r|^{-2\beta /\nu }\,g\left( \frac{r}{t^{1/z_{c}}}\right)
\\
C(t,s) &\approx &s^{-2\beta /\nu z_{c}}\,f_{C}\left( \frac{t}{s}\right) .
\end{eqnarray*}
For the 1D Ising model at zero temperature we have
\begin{eqnarray}
C_{r}(t) &\approx &\mathrm{erfc}\left( \frac{|r|}{2t^{1/2}}\right)
\nonumber \\
C(t,s) &\approx &\frac{2}{\pi }\arctan \left( \frac{2s}{t-s}\right) ^{\frac{1%
}{2}}.
\label{Cts_Is1D}
\end{eqnarray}
The latter formulas are compatible with the former ones,
taking into account that $\beta =0$ for the 1D Ising model.
Otherwise stated, the absence of an
anomalous dimension implies that $C(t,s)$ is not small in the
critical region, in contrast to the generic cases considered in the previous
section.

From (\ref{Cts_Is1D}) we obtain
\[
f_{C^{\prime }}(x)=\frac{x}{\pi (x+1)}\sqrt{\frac{2}{x-1}}.
\]

The critical temperature $T_{c}$ being equal to zero, we define the
dimensionless response function
\[
\tilde{R}(t,s)=T\frac{\delta \langle \sigma (t)\rangle }{\delta H(s)}.
\]
In the scaling region ($1\ll s\sim t$), this function is found to behave as
\[
\tilde{R}(t,s)\approx s^{-1}\,f_{\tilde{R}}\left( \frac{t}{s}\right) ,
\]
where
\[
f_{\tilde{R}}(x)=\frac{1}{\pi \sqrt{2(x-1)}}.
\]
This again is compatible with the generic case, with $\beta =0$.

The reduced magnetisation $\rho _{\TRM}(t,s)$ and
the fluctuation-dissipation ratio $X(t,s)$ can be computed explicitly.
Both quantities only depend on $t/s$, or equivalently on $C$,
in the scaling regime.
One finds, in this regime \cite{gl1D,zann},
\[
\rho _{\TRM}(C)=\frac{\sqrt{2}}{\pi }\arctan \left( \frac{1}{\sqrt{2}}\tan
\frac{\pi C}{2}\right) ,
\]
while $X$ is more simply written in terms of the ratio $x=t/s$ as
\[
X(t,s)=\frac{f_{\tilde{R}}(x)}{f_{C^{\prime }}(x)}=\frac{x+1}{2x}.
\]
We note once again that the fact that $\beta =0$ implies no dependence in $s$
in these quantities.
Finally, the last equation implies for the limiting ratio
\[
X_{\infty }=\lim_{s\rightarrow \infty }\lim_{t\rightarrow \infty }X(t,s)=%
\frac{1}{2}.
\]

\section{\ Discussion}

At criticality, for the generic cases of the spherical model and
of the 2D Ising model, $X(t,s)$ is not a function of $C(t,s)$.
It is instead a function of the ratio $x=t/s$,
or equivalently of $s^{2\beta/\nu z_c}C(t,s)=f_C(x)$.
In this last representation, the value of $X$ at the origin is equal
to $X_\infty$.
Then the fluctuation-dissipation ratio increases
and reaches the limit value 1 when
the abscissa $f_C(x)$ goes to infinity, that is, for $x\to 1$,
where the fluctuation-dissipation theorem holds.

Is the amplitude ratio $X_\infty$ related to equilibrium quantities?
This remains an interesting open question.
More generally, do the above results on the fluctuation-dissipation ratio
admit a static interpretation,
e.g. in terms of the distribution of overlaps $P(q)$ \cite{franz}?
Strictly speaking, the existence of a non-trivial $X_\infty$
should imply the presence of an unexpected discrete component in $P(q)$.
We mention a recent work on related matters \cite{bert},
where the finite-size behaviour of $P(q)$ for the 2D X-Y model
is related to the finite-time behaviour of $\rho(t,s)$.

A recent analysis \cite{henkel}, based on conformal invariance, predicts the
following form of the response function
\begin{equation}
R(t,s)=r_{0}(t-s)^{-A}\left( \frac{t}{s}\right) ^{-B},
\label{malte}
\end{equation}
without predicting the values of exponents appearing in the right side.
This prediction should hold for a large class of systems.
We note in particular that, for the spherical model,
equations (\ref{Rts_coars}) and (\ref{Rts_crit}),
together with the explicit forms
of the scaling functions $f_R(x)$, given in sections 2 and 3,
confirm this prediction,
which is also verified by numerical computations
on the Ising model in two and three dimensions~\cite{henkel}.
The analytical results for the 1D Ising model given in section 4
do not, however, satisfy the prediction (\ref{malte}).

Finally it is worth adding a few words on the comparison between the results
reviewed here and those reviewed in ref. \cite{glreview} for urn models.
For the zeta urn model, the situation at criticality is
in all aspects similar to
that of a generic ferromagnetic model, as described in section 3.
However the prediction (\ref{malte}) is not fulfilled by this model.
In the low-temperature phase, the results obtained for the zeta urn model
do not fall in the framework reviewed in section 2, valid
for a coarsening system.
Finally, the results obtained for the backgammon model at $T=0$
are rather different from the generic behaviour of a ferromagnetic model.
A natural explanation of the discrepancy between urn models
and ferromagnetic models is that in the former case
the system is rather subject to condensation than to coarsening.

%
%
%
%


\begin{table}[htbp]
\caption{
Static and dynamical exponents of the ferromagnetic spherical model,
and of the Ising model in one and two dimensions.
First group: usual static critical exponents $\eta$, $\beta$, and $\nu$
(equilibrium).
Second group: zero-temperature dynamical exponents~$z$ and $\lambda$
(coarsening below $T_c$).
Third group: dynamic critical exponents $z_c$, $\lambda_c$, and $\Theta_c$
(nonequilibrium critical dynamics).}
\begin{center}
\begin{tabular}{|c|c|c|c|c|}
\hline
exponent&spherical $(2<D<4)$&spherical $(D>4)$& 2D Ising &1D Ising \\
\hline
$\eta$&$0$&$0$&$1/4$&1\\
$\beta$&$1/2$&$1/2$&$1/8$&0\\
$\nu$&$1/(D-2)$&$1/2$&$1$&1\\
\hline
$z$&$2$&$2$&$2$&\\
$\lambda$&$D/2$&$D/2$&$\approx1.25$&\\
\hline
$z_c$&$2$&$2$&$\approx2.17$&2\\
$\lambda_c$&$3D/2-2$&$D$&$\approx1.59$&1\\
$\Theta_c$&$1-D/4$&$0$&$\approx0.19$&0\\
\hline
\end{tabular}
\end{center}
\end{table}


\end{document}